\documentclass[runningheads]{llncs}

\usepackage[T1]{fontenc}

\usepackage[table,xcdraw,dvipsnames]{xcolor}

\usepackage{cite}
\usepackage{amsmath}
\usepackage{graphicx,calc}
\usepackage{xspace}
\usepackage{tikz}
    \usetikzlibrary{arrows}

\usepackage{booktabs}

\usepackage{subcaption}
\usepackage{hyperref}
\usepackage{wrapfig}

\usepackage[ruled, linesnumbered]{algorithm2e}

\SetCommentSty{mycommfont}
\SetKwRepeat{Do}{do}{while}

\newcommand{\tool}[1]{\textsc{#1}\xspace}

\newcommand{\racerF}{\tool{RacerF}}

\newcommand{\goblint}{\tool{Goblint}}

\newcommand{\true}{\textsf{true}\xspace}
\newcommand{\false}{\textsf{false}\xspace}
\newcommand{\unknown}{\textsf{unknown}\xspace}

\begin{document}

\title{\racerF: Data Race Detection with Frama-C}
\subtitle{(Competition Contribution)\vspace*{-8mm}}

\renewcommand{\orcidID}[1]{{\href{https://orcid.org/#1}{\protect\raisebox{3.25pt}{\protect\includegraphics{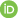}}}}}

\author{Tomáš Dacík\thanks{Jury member.}\inst{1}\orcidID{0000-0003-4083-8943}
\and Tomáš Vojnar\inst{1,2}\orcidID{0000-0002-2746-8792}\vspace*{-2mm}}
\authorrunning{T. Dacík and T. Vojnar}
\institute{ Faculty of Information Technology, Brno University of Technology,
Czech Republic \and Faculty of Informatics, Masaryk University, Czech Republic }

\maketitle

\begin{abstract} \vspace*{-3mm} \racerF is a static analyser for detection of
data races in multithreaded C programs implemented as a plugin of the Frama-C
platform. The approach behind \racerF is mostly heuristic and relies on analysis
of the sequential behaviour of particular threads whose results are generalised
using a combination of under- and over-approximating techniques to allow
analysis of the multithreading behaviour. In particular, in SV-COMP'25, \racerF
relies on the Frama-C's abstract interpreter \textsc{EVA} to perform the
analysis of the sequential behaviour. Although \racerF does not provide any
formal guarantees, it ranked second in the \textit{NoDataRace-Main}
sub-category, providing the largest number of correct results (when excluding
metaverifiers) and just 4 false positives.
%
%
\end{abstract}

\vspace*{-6mm} \section{Verification Approach}  \label{sec:overview} \vspace*{-2mm}

\racerF is a static analyser whose primary goal is to provide fast and scalable
data race detection, without sacrificing too much precision. It relies on using
a backend static analyser that summarises (in a way specific for the chosen
analyser) the sequential behaviour of particular threads -- or, more precisely,
classes of threads since \mbox{\racerF} distinguishes threads 
%
%
according to their entry point function only. The existing threads are
discovered incrementally since a newly discovered thread may create further
threads. The results of the analysis of the sequential behaviour are generalised
for the multi-threaded setting using a combination of an under-approximating and
an over-approximating strategy.

\racerF comes with several analysis backends differing in their scalability and
precision. In SV-COMP'25, we use the most precise of them -- Frama-C's value
analysis plugin EVA  based on abstract interpretation combining several abstract
domains 
%
%
to characterize sets of integers 
and addresses that may get assigned to
the program variables~\cite{EVA}. Over the backend's results, \racerF runs
several analyses  
%
%
to perform data race detection as a final step. We briefly describe these
analyses in the rest of this section. More details can be found in
\cite{RacerF}. 

\enlargethispage{6mm}

\vspace{-2mm} \subsection{Analysing Multithreaded Programs Using 
Sequential Behaviour} 

The main idea behind \racerF is to discover all program threads, distinguished
statically according to their entry functions, and analyse them as sequential
programs, starting with initial states that are computed using a set of
equations over currently discovered threads. We have two strategies to construct
those equations, based on under-approximating or over-approximating thread
behaviours. The approximations computed by the different analysis backends are
then available to subsequent analysis phases performed by \racerF, which may but
need not respect the guaranteed under-/over-approximation.

\paragraph{Under-approximating strategy.}

The idea behind this strategy is to ignore all interleavings. Each thread is
analysed with the initial state corresponding to the join of states discovered
as reachable at its create statements, and its writes to global variables are
not propagated to other threads. When there is no cyclic dependency in thread creation, it is
enough to analyse each thread just once.

\vspace{-2.5mm} \paragraph{Over-approximating strategy.}

This strategy over-approximates considered interleavings (including also those
ruled out by some synchronization method). In this setting, each thread is
analysed with the initial state given as the join of all states encountered by
analyses of all discovered threads. This strategy may require several
re-analyses of individual threads, but we aggressively use widening to reduce
this number. This strategy needs to be accompanied by a program transformation
that ensures that the computed over-approximation of the initial state is not
lost during the sequential analysis of threads. Intuitively, the exchange of
information between thread analyses happens at the initial states only, and so
we need to preserve the information obtained from analyses of other threads
throughout the re-analysis of a given thread and not destroy it by some local
assignments before it gets propagated to a code location where it actually
matters (see \cite{RacerF} for more details). This transformation was
implemented after our submission to SV-COMP'25, but its absence does not
manifest in the results (it would manifest only on programs that our memory
access analysis cannot handle precisely at that time.)


\vspace{-2.5mm} \paragraph{Combined strategy.}

For SV-COMP'25, we provide a wrapper that combines both strategies in a natural
way. The under-approximation is only allowed to report races, and the
over-approximation is only allowed to claim a program race-free. We start with
the over-approximation, and if it is inconclusive, we run the
under-approximation. If both strategies are inconclusive, we report
\unknown. The situation in which running the combination is beneficial
within SV-COMP is, however, quite rare as the result of using purely under- or
over-approximation would lead to results different from using their combination
in 11 out of 1029 data race benchmarks only.


\enlargethispage{8mm}

\vspace{-3mm} 
\subsection{Analysis Pipeline} 

We call an \emph{analysis context} a triple consisting of a thread, a sequence
of a fixed number of the latest calls (each of them formed by a call site and
the function called, with the number being a parameter of the analysis -- for
SV-COMP'25, set to two), and a program statement. We run the following pipeline
of analyses to detect data races.

\emph{Lockset analysis.} This analysis computes the set of sets of locks held
for each program context (using an algorithm inspired by \cite{racerx}) and provides must- and may-queries to check whether two
contexts are guarded by a common lock. Trylocks and read-write locks are
supported.

\emph{Active-threads analysis.} The goal of this analysis is to provide
information which threads may- and must-run in parallel in particular contexts.
This information is then used to not report data races on memory accesses where
one happens before the thread of the second is created, or after it is surely
joined.
    
\emph{Memory access analysis.} EVA represents memory addresses as pairs
consisting of a \emph{base} and an \emph{offset}. Each variable (global, local,
or formal) has its associated base. Dynamic allocations have so-called
\textit{dynamic bases}, either strong or weak (representing multiple
allocations). The memory access analysis tracks accesses to memory bases and
distinguishes their states in a way inspired by~\cite{Eraser}. Those states are used to select bases for which data race detection should be performed.

    
\emph{Data race detection.} Finally, over the discovered memory accesses, a may and a must data race detection is performed. Intuitively,
the may date race detection checks whether at least one of the accesses is a
write access, the accesses may come from distinct threads and may happen in
parallel, the concerned locksets may have empty intersection, and the concerned
memory blocks may overlap (with the must data race detection modified
accordingly and accompanied by several more checks to eliminate common sources of false positives).

\vspace{-2.5mm} \paragraph{Reducing Incorrect Verdicts.} 

As already mentioned, to reduce the number of incorrect verdicts, we can leverage
the distinction of the detected races to may and must races, essentially
reporting a data race only when a must data race is detected, and reporting no
data race only if no may data race is found.
Otherwise, an \unknown result is produced.
%

Further, since the SV-COMP'25 benchmark also contains programs implementing
lock-free algorithms, which we cannot analyse, we have implemented a simple
detection of the code pattern of \emph{active waiting} (essentially detecting
empty control loops), which is the common denominator of such programs, and
report \unknown if we encounter such active waiting.

\begin{table*}[!t]
    \vspace{-4mm}
    \centering
    \caption{Results of \racerF in \textit{NoDataRace-Main} sub-category.}
    \label{table:svcomp}
    \begin{tabular}{@{}lrrrrrrrrrr@{}}
\toprule
& \multicolumn{2}{c}{Correct} & & \multicolumn{2}{c}{Wrong} & & \multicolumn{2}{c}{Unconfirmed}\\

\cmidrule{2-3} \cmidrule{5-6} \cmidrule{8-9}
Analyser$\hspace{30pt}$ & $\;$True & $\;$False & $\;\;\;\;$ & $\;$True & $\;\;\;\;$False &$\;\;\;\;\;$& True &\;\;\; False & $\;\;\;\;$ Score &\;\;\;\;Time [s]\\
\midrule

\textsc{RacerF} & 674 & 68 & & 0 & 4 & &0 & 30 & 1352 & 1300\\
\bottomrule
\end{tabular}
    \vspace{-6mm}
\end{table*}

\vspace{-3mm} \section{Strengths and Weaknesses} \vspace{-2mm}
\label{sec:strenghts-and-weakneses}

Despite its heuristic nature, \racerF can provide very precise results as
Table~\ref{table:svcomp} shows \cite{SVCOMP25}. \racerF does not provide a
\true/\false verdict for 253 programs, of which 17 cannot be
parsed by Frama-C, 60 contain unsupported features (3 cases of semaphores, 57
cases of custom atomic functions), and in 176 cases, our analyser returns
\unknown (in 20 cases, active waiting is detected; and in 156 cases, our
may/must data race classification is inconclusive).

\enlargethispage{2mm}

\vspace{-2.5mm} \paragraph{Strengths.}

\racerF is quite fast. In just 8 cases, it needs more than 5 seconds to compute
a result, and it also never runs out of time or memory. This is in stark
contrast with model checkers. The only competitive SV-COMP'25 participant from
this point of view is \goblint \cite{goblint, GOBLINT-SVCOMP24}, which, however, also timed out in a few cases. What we are especially proud of is that \racerF is the only SV-COMP'25
participant (excluding tools with a negative score) which provides correct results for all 5 programs derived from the Linux kernel in the
\texttt{ldv-linux-3.14-races} benchmark. Besides it, just \goblint is able to
provide the correct result for one of those tasks. All the other tools either
run out of resources, fail, or return an \unknown verdict. \racerF needs just 2
minutes for one of them and about 30 seconds for the others.


\vspace{-2.5mm} \paragraph{Weaknesses.}

\racerF does not provide any formal guarantees for its results. There are 4
cases in which our heuristic approach reports a false alarm. All of them rule
out a data race by some intricate condition that is out of the scope of our
analyses. Two of them (\texttt{pthread-ext/09\_fmaxsym-zero.c} and
\texttt{pthread-ext/11\_fmaxsymopt-zero.c}) are variants of the same
program that relies on the fact that an array is zero-initialized and a certain
condition cannot hold more than once throughout the run of the program. The
other two \mbox{programs \texttt{goblint-regression/13-privatized\_40-traces-ex-6\_true.c}} and \texttt{pthread\!-\!atomic/time\_var\_mutex.c} use synchronisation
patterns that are beyond the scope of our lockset analysis.

Another weakness of our analyser is a lack of path-sensitivity. In some cases,
especially on the \texttt{ldv-races} benchmark, \racerF returns \unknown because
it considers some trivially infeasible path that prevents it from classifying
the encountered issue as a must data race. In future versions, this can be
improved by adding preprocessing to eliminate trivial branching conditions and
dead code. 

Finally, \racerF can only generate trivial witnesses and thus almost one third
of its \false answers is unconfirmed. In fact, \racerF provides
traces for detected races, but we have not yet implemented their conversion
to witnesses.

\vspace{-3mm} \section{Software Project and Tool Setup} \vspace{-1mm}

\racerF is implemented in OCaml as a plugin of the Frama-C platform
\cite{Frama-C1, Frama-C2}. It relies on the Frama-C's fork of CIL \cite{CIL} as
its frontend and uses the abstract interpretation plugin EVA \cite{EVA} to
perform sequential analyses of individual threads. The library \mbox{OCamlgraph}
\cite{ocamlgraph} is used to represent several graph structures and solve systems
of equations created during thread analysis. 
In SV-COMP'25, \racerF is run via a Python script \texttt{racerf-sv.py} which
performs the following:

\emph{Preprocessing}. Since Frama-C discards the $\mathtt{\_Atomic}$ attributes
during parsing, we encode them as type aliases, e.g., the declaration
\mbox{\texttt{\_Atomic\;int\;x}} is transformed to \mbox{\texttt{atomic\_int\;x}}.
We also further preprocess several aspects of input that are
problematic for Frama-C (e.g. we replace variable-length arrays in non-final
fields of structures with arrays of a constant size). The preprocessing is
implemented in the script \texttt{preprocess.py}.
    
\emph{Combination of under- and over-approximation.} The script runs two
configurations of \racerF and reports the verdict as discussed in Section
\ref{sec:overview}.

\emph{Witness postprocessing.} During the competition, it turned out that \mbox{\racerF} produces syntactically incorrect witnesses. The script
\texttt{fix\_witness.py} fixes this by filling in the correct metadata.
The witness graph
itself is not modified.  

\enlargethispage{6mm}


\paragraph{Usage.} A binary of \racerF is available at Zenodo \cite{racerF-zenodo} (the only runtime dependencies are \textsc{GCC} and Python version 3.10 or newer). The wrapper script can be run as (see the attached file \texttt{README.md} for more information):\vspace*{-1.2mm}

\begin{center}
  \texttt{./racerf-sv.py -machdep=[gcc\_x86\_32|gcc\_x86\_64] input.c}
\end{center}

\paragraph{Software project.}

\racerF is available under the MIT license as a part of the
authors' project of concurrency analysers for C, maintained by Tomáš Dacík. 

\paragraph{Participation.} \racerF participated in the \textit{NoDataRace-Main} sub-category only.


\newlength\myheight
\newlength\mydepth
\settototalheight\myheight{Xygp}
\settodepth\mydepth{Xygp}
\setlength\fboxsep{0pt}
\newcommand*\inlinegraphics[1]{%
  \settototalheight\myheight{Xygp}%
  \settodepth\mydepth{Xygp}%
  \raisebox{-\mydepth}{\includegraphics[height=\myheight]{#1}}%
}

\paragraph{Data-Availability Statement.} 
\racerF is available under the MIT license at \url{https://github.com/TDacik/Deadlock_Racer}.
The version participating in SV-COMP'25 is available under the tag \textsf{svcomp25} and archived at Zenodo \cite{racerF-zenodo}.

\paragraph{Acknowledgements.} We appreciate discussions with Julien Signoles
from the Frama-C team at CEA, France and with Adam Rogalewicz from FIT BUT,
Czechia. The research was supported by the Czech Science Foundation project
23-06506S and the FIT BUT internal project FIT-S-23-8151. Tomáš Dacík was
supported by the Brno PhD Talent Scholarship of the Brno City Municipality. The
collaboration with Julien Signoles was supported by the project VASSAL:
``Verification and Analysis for Safety and Security of Applications in Life''
funded by the European Union under Horizon Europe WIDERA Coordination and
Support Action/Grant Agreement No. 101160022. \raisebox{0.5mm}{\inlinegraphics{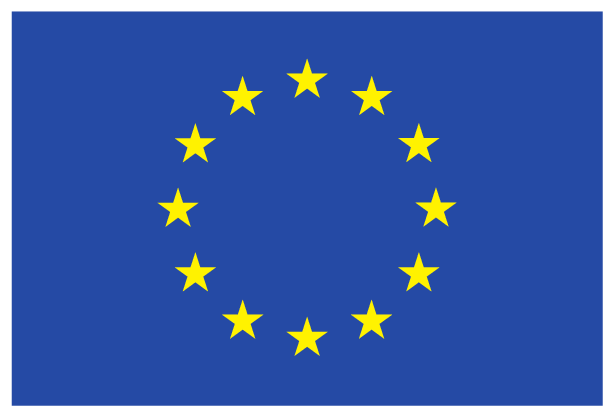}}

\enlargethispage{8mm}


\vspace{-2mm}

\bibliographystyle{splncs04}
\bibliography{references.bib}

\end{document}